\documentclass[usenatbib]{mn2e}

\usepackage[dvips]{graphicx}

\title[Nuclear activity and stellar mass in galaxies]{The Relation between Nuclear Activity and Stellar Mass in Galaxies}
\author[G. La Mura et al.]{G. La Mura,$^{1,2}$\footnotemark D. Bindoni,$^1$ S. Ciroi,$^{1,2}$ V. Cracco,$^{1,2}$ R. D'Abrusco,$^3$ \newauthor P. Rafanelli,$^{1}$ L. Vaona$^1$ \\
$^1$Department of Physics and Astronomy - University of Padua, Vicolo dell'Osservatorio 3, 35122, Padova, Italy \\
$^2$INAF - Astronomical Observatory of Padua, Vicolo dell'Osservatorio 5, 35122, Padova, Italy \\
$^3$Harvard - Smithsonian Astrophysical Observatory, 60 Garden Street, Cambridge, MA 02138}

\fnsymbol{footnote}
\begin{document}
\newcommand{\pind}{{\it P}-index}
\newcommand{\ion}[2]{#1~{\small #2}}

\maketitle
\begin{abstract}
The existence of correlations between nuclear properties of galaxies, such as the mass of their central black holes, and larger scale features, like the bulge mass and luminosity, represent a fundamental constraint on galaxy evolution. Although the actual reasons for these relations have not yet been identified, it is widely believed that they could stem from a connection between the processes that lead to black hole growth and stellar mass assembly. The problem of understanding how the processes of nuclear activity and star formation can affect each other became known to the literature as the Starburst-AGN connection. Despite years of investigation, the physical mechanisms which lie at the basis of this relation are known only in part. In this work, we analyze the problem of star formation and nuclear activity in a large sample of galaxies. We study the relations between the properties of the nuclear environments and of their host galaxies. We find that the mass of the stellar component within the galaxies of our sample is a critical parameter, that we have to consider in an evolutionary sequence, which provides further insight in the connection between AGN and star formation processes.
\end{abstract}

\begin{keywords}
galaxies: active -- galaxies: evolution -- galaxies: ISM -- galaxies: nuclei -- galaxies: stellar content
\end{keywords}

\section{Introduction}
\footnotetext{$^\star$E-mail: giovanni.lamura@unipd.it}
Recent statistical investigations pointed out that the nuclear regions of approximately 20 per cent of nearby galaxies show emission from ionized gas \citep{Taniguchi03}. The presence of this component is mostly explained by processes related either to intense star formation or to non-thermal activity in the nucleus. In principle, it is possible to distinguish the signature of the two different possibilities from the spectrum emitted by the ionized gas \citep[see e. g.][]{Veilleux87, Kewley06}. From a practical point of view, however, the task is made difficult because of the trend, observed in large samples of sources, to settle on to a smooth sequence, rather than forming distinct distributions. The problem is partially due to the difficulties related to the spectroscopic measurements, which lie at the basis of the technique, but the existence of a population of transition objects, characterized by intermediate properties, also plays an important role.

There are in fact numerous works where the relationship between nuclear activity and star formation processes in the central regions of galaxies is addressed, mainly taking into account Seyfert 2 galaxies \citep*[e. g.][]{Ivanov00, Gonzalez01, Gu01, Joguet01, Storchi-Bergmann01, CidFernandes04}. The properties of Type 1 objects are clearly harder to understand, because the light, coming directly from the nuclear source, usually outshines the surrounding regions. This problem leaves open the question whether the sources are intrinsically different or not, which is a concern both for black hole--host galaxy relationships and for AGN unification theories. At present, evidence for recent nuclear star formation exists also for Type 1 AGNs \citep[e. g.][]{Davies07} and it is believed that there is little or no difference with respect to Type 2 objects under this point of view, but the debate is still in progress.

As a result, though some kind of connection between nuclear activity and star formation must exist, its physical role is not yet clarified. Several theories were developed, mainly invoking effects that could lead from nuclear star formation to AGNs and then evaluating the energetic feedback on the surrounding medium \citep{Weedman83, Norman88, Scoville88, Heckman89, Taniguchi99, Ebisuzaki01, Mouri02}. However, the possibility that the relation may change, depending on the host galaxy properties, should also be taken into account. Indeed, it has been pointed out that the high velocity outflows, produced by the most massive stars of a young stellar population, are very unlikely to feed nuclear activity and they can even sweep the interstellar medium of low mass galaxies, preventing the onset of further star formation  or nuclear activity. In very massive galaxies, on the other hand, the large amount of gas trapped in the gravitational field could serve as fuel for a powerful AGN, which is again expected to stop the formation of new stars \citep[and references therein]{Monaco05, Feruglio10}. Such triggers may explain the reason why star formation is currently observed over a typically limited range of medium mass galaxies.

In this work, we study the properties of the circum-nuclear regions of galaxies, whose spectra indicate the existence of an interstellar medium (ISM), which is ionized either by hot young stars or by non-thermal activity of the nucleus. After selecting our targets, we carry out an extensive analysis of their stellar populations, their chemical compositions, and their ionization sources. Comparing such properties with those of quiescent galaxies, we find additional evidence of close relationships between nuclear activity and star formation processes, together with several sequences, indicating a possible evolutionary connection between the different classes.

Our work is organized as follows: in \S2 we describe the selection of our sample and its basilar properties; \S3 introduces the data analysis and the strategies that we applied to perform our measurements; in \S4 we present a discussion of our results and of their interpretation, while our conclusions are summarized in \S5.

\section{Sample selection and description}
To investigate the relationship between star forming galaxies (SFG) and AGNs, our work required the selection of a wide sample of emission line galaxies. The Sloan Digital Sky Survey Data Release 7 \citep[SDSS DR7, see][]{Abazajian09} provided the ideal archive of spectroscopic data for our purposes. Here we looked for the spectra of galaxies with ionized gas components in the nucleus. The classification of these objects according to their ionization source led to the introduction of a diagnostic technique, which could successfully manage line emitting sources, including both Type 1 and Type 2 AGNs. Our choice fell on the diagram plotting $\log(F_{\rm [O II] 3727} / F_{\rm [O III] 5007})$ against $\log(F_{\rm [O I] 6300} / F_{\rm [O III] 5007})$ (hereafter the $O_{123}$ diagram). This instrument is a combination of two diagnostic ratios that express, respectively, the ionization degree and the hardness of the ionizing spectrum at large radii from the source \citep{Kewley06, Shields90}. It provides the clear advantages of being compatible with all types of AGN spectra, to be nearly independent from the chemical composition of gas, and to be free from stellar absorption contaminations, which are relevant for most recombination lines. On the other hand, it may suffer from differential extinction effects, while the [\ion{O}{I}] emission line is often weak and difficult to measure in noisy spectra. For this reason we collected the spectra of galaxies with detected emission lines of [\ion{O}{III}]$\,\lambda 5007$, [\ion{O}{II}]$\,\lambda\lambda 3727,3729$ and [\ion{O}{I}]$\,\lambda 6300$, requiring a S/N~$\geq 3$ for [\ion{O}{I}]. We extracted 119226 spectra that we plot in the $O_{123}$ diagram of Fig.~1.

\begin{figure}
\begin{center}
\includegraphics[width=8.4cm]{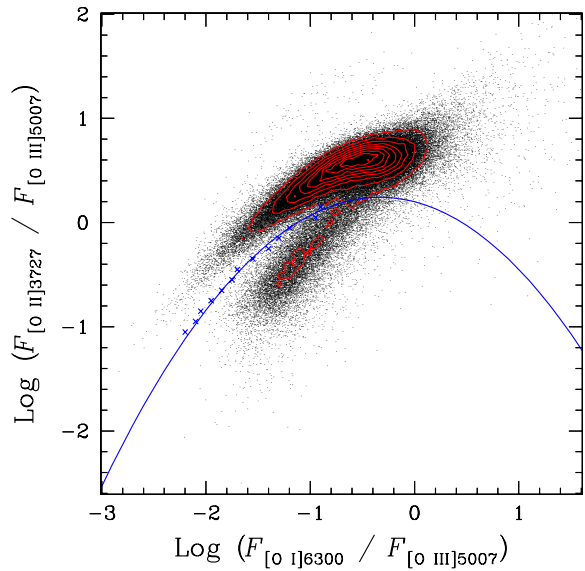}
\end{center}
\caption{The sample of 119223 emission line galaxies plotted on the $O_{123}$ diagnostic diagram. The red contours show the density distribution of sources in levels spaced by 100 objects per 0.05~dex square bin on the diagram. The blue crosses mark the bins corresponding to local minima in the distribution, while the blue curve is the borderline defined in Eq.~(1), below which 90 per cent of the objects are AGN powered sources.\label{f01}}
\end{figure}
Comparing the $O_{123}$ diagnostic diagram with the Veilleux-Osterbrock (VO) classification, updated by \citet{Kewley06}, it can be realized that the observed sequences are generally populated by different sources: the upper sequence is mostly consisting of star forming galaxies, the lower of active nuclei, and the overlapping region of LINERs and transition sources. Subdividing the plot in bins of 0.1~dex in the range $-1.2 \leq \log(F_{\rm [O II] 3727} / F_{\rm [O III] 5007}) \leq 0.3$, where the separation between the sequences is clearer, we computed the number of sources within 0.1~dex wide bins of $\log(F_{\rm [O I] 6300} / F_{\rm [O III] 5007})$. We, thus, built histograms showing the two-peaked source distributions and we located the minimum for each bin. The locations of such minimum points were subsequently interpolated by a polynomial function, derived by means of a least squares regression, in the form of:
$$\log \left(\frac{F_{\rm [O\, II]\, 3727}}{F_{\rm [O\, III]\, 5007}}\right) = 0.20 - 0.25\, \log \left(\frac{F_{\rm [O\, I]\, 6300}}{F_{\rm [O\, III]\, 5007}}\right) -$$
$$-0.39\, \left[\log \left(\frac{F_{\rm [O\, I]\, 6300}}{F_{\rm [O\, III]\, 5007}}\right)\right]^2. \eqno(1)$$
\begin{figure*}
\begin{center}
\includegraphics[width=5.8cm]{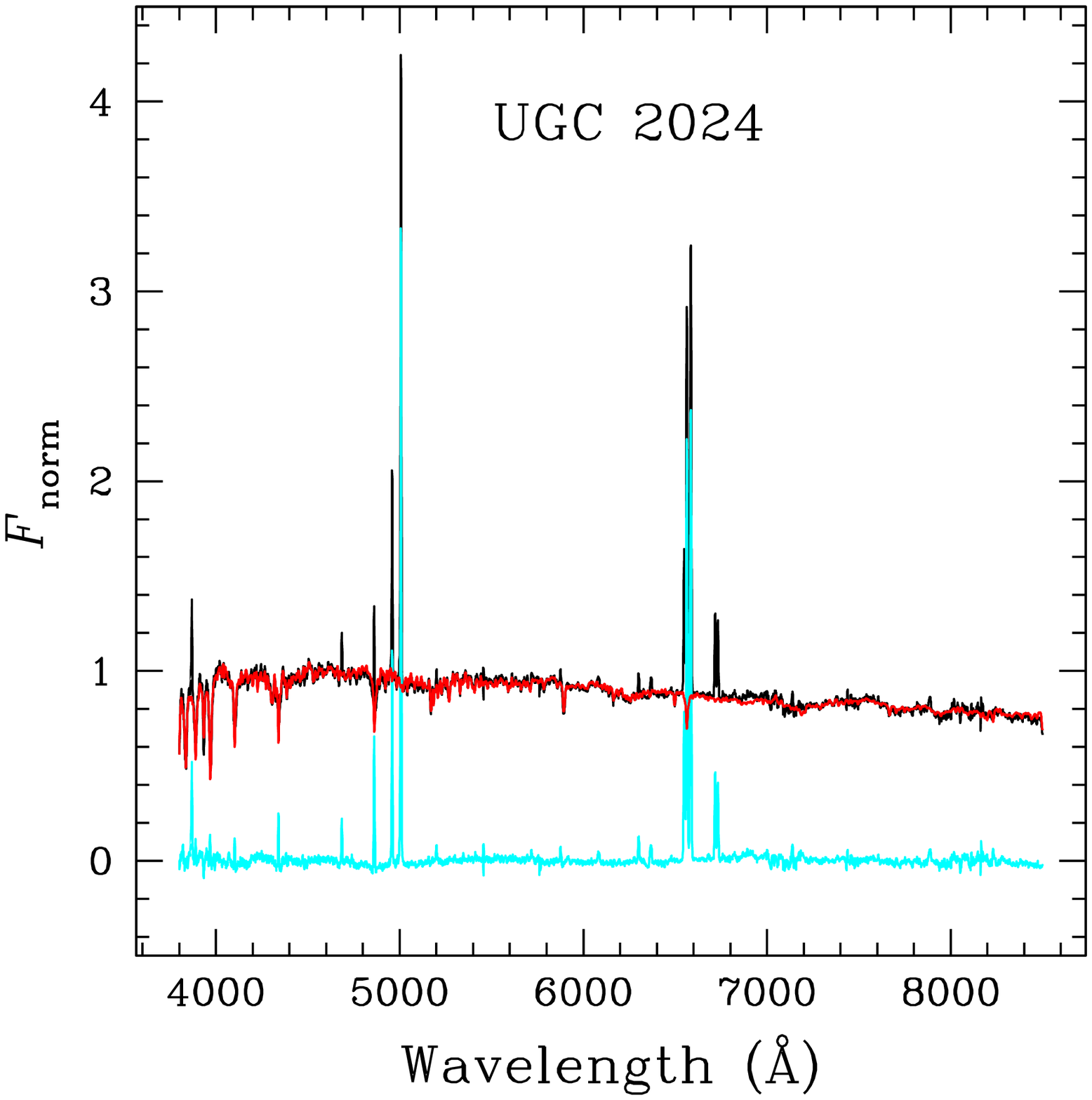}
\includegraphics[width=5.8cm]{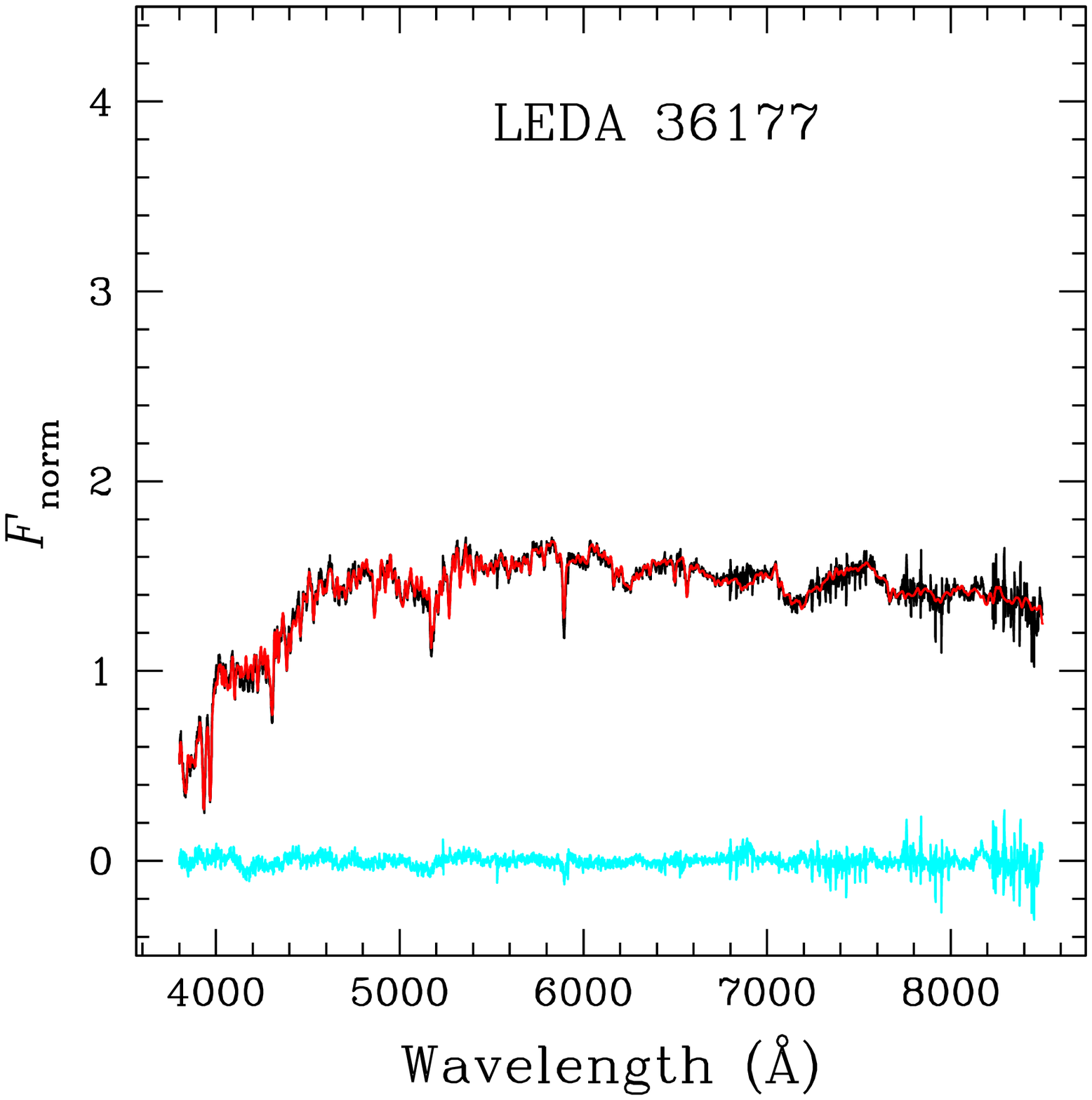}
\includegraphics[width=5.8cm]{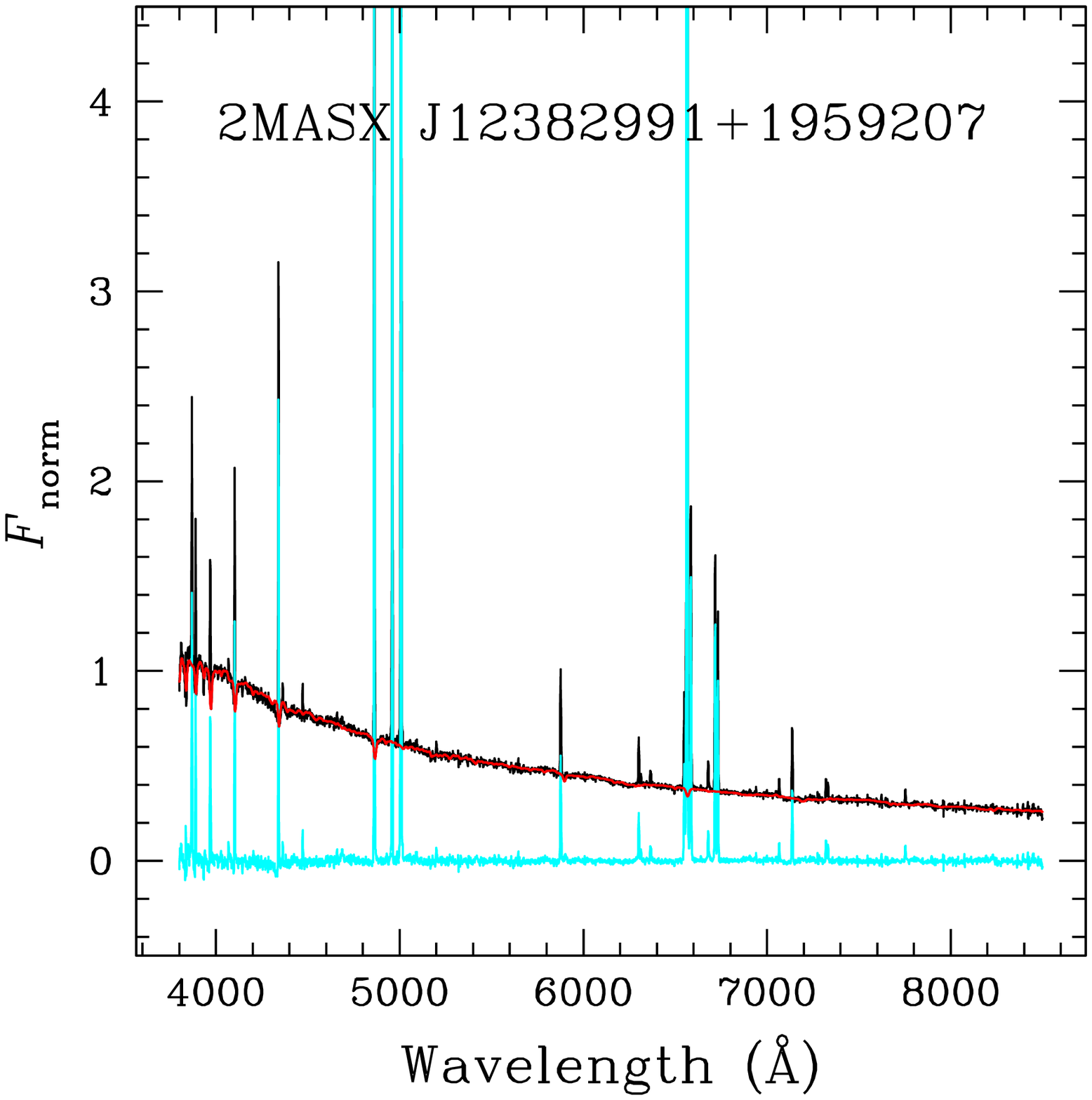}
\end{center}
\caption{Examples of stellar continuum spectral synthesis for the case of a Seyfert 2 galaxy (left panel), a normal galaxy (central panel) and a star forming galaxy (right panel). The continua were normalized with respect to the flux at 4100~\AA\ and the plots show the observed spectra in black, the synthetic stellar continua in red and the fitting residuals in cyan. We can easily note the Balmer line absorption features that may significantly affect the ratio of important diagnostic lines, especially in the case of S2Gs.}
\end{figure*}
By taking into account only the objects below the curve, we isolated a sample of approximately 16000 objects that consists of 60 per cent Seyfert galaxies, including both broad and narrow line sources, 31 per cent composite Seyfert - \ion{H}{II} galaxies, 4 per cent LINERs, and 5 per cent SFGs.

A detailed classification of the narrow line emitting objects, identified by means of a line profile fitting code described in a companion paper \citep[GGFIT,][]{Vaona12}, can be performed by application of the classic VO diagnostic diagrams. Before proceeding, however, we have to account for the stellar contributions, which significantly affect the intensities of recombination lines like H$\alpha$ and H$\beta$. These contributions were estimated with a spectral synthesis technique, based on the {\small STARLIGHT} code \citep{CidFernandes05}, using a set of 92 simple stellar population spectra (SSP), obtained combining 23 ages (from $10^6$~yr up to $13 \cdot 10^9$~yr) and 4 metallicities (Z = 0.004, 0.008, 0.020, and 0.050) from the library of \citet{Bruzual03}. In order to fit the observed spectra, we had to correct them for the extinction arising in our own Galaxy, according to the extinction map provided by the {\small NED} service\footnotemark \footnotetext{Nasa Extragalactic Database, addressed at \\ {\tt http://nedwww.ipac.caltech.edu}}, to remove the cosmological redshift, and to re-sample the data in a fixed wavelength dispersion. We performed these operations by means of the {\small IRAF} tasks {\tt deredden}, {\tt newredshift}, and {\tt dispcor}. Such procedures were meant to derive accurate models of the spectral continua and absorption lines, thus producing residuals of pure emission line spectra, provided that the $S/N$ of original data was high enough. A threshold value of $S/N \geq 10$ at 5500~\AA\ granted the possibility to apply the fitting procedures with reliable results.

Since we are interested in the properties of the circum-nuclear regions of the galaxies, we restrict the sample to objects in the redshift range of $0.04 \leq z \leq 0.08$, where the SDSS fiber aperture covers a region of $2.4 - 4.8\,$kpc in radius. In this study we considered only Seyfert 2 galaxies (S2G), collecting 2153 objects. A set of 1302 SFGs was identified on the diagnostic diagrams, by applying the constraints \citep{Rafanelli09}:
$$\log ([{\rm O\, III}] / {\rm H}\beta) < 0.61 / [\log ([{\rm N\, II}] / {\rm H}\alpha)] + 1.3 \eqno(2a)$$
$$\log ([{\rm O\, III}] / {\rm H}\beta) < 0.72 / [\log ([{\rm S\, II}] / {\rm H}\alpha)] + 1.3 \eqno(2b)$$
$$\log ([{\rm O\, III}] / {\rm H}\beta) < 0.73 / [\log ([{\rm O\, I}] / {\rm H}\alpha)] + 1.33 \eqno(2c)$$
We finally collected an additional set of galaxy spectra, without detected emission lines, selecting 3000 objects, above the adopted $S / N > 10$ threshold at 5500~\AA, that were observed in the same redshift range. These formed our Normal Galaxy sample (NG), which we compared to the line emitting objects. More general properties of the AGN class, including broad line emitting sources, will be dealt with in \citet{Vaona12}.

\section{Data analysis}
\subsection{Stellar populations}
The circum-nuclear stellar populations of our galaxy sample contribute to the observed spectra with various components. Depending on the object distance, different fractions of the target galaxy lie in the region covered by the spectrograph fibers. As a result, we observe the composite spectrum of integrated stellar populations, with their characteristic continuum and absorption lines, overlapped to the emission lines, produced by the diffuse ionized gas. To apply the classification procedures, described above, and to investigate the physical properties of the sample, we need to distinguish between these components. For this purpose, we reproduced the stellar populations, which originate the spectra of all the selected galaxies, using {\small STARLIGHT}. The models were computed applying linear combinations of the 92 synthetic spectra of SSPs, with homogeneous age and chemical composition, extracted from the library of \citet{Bruzual03} and sampled at the same spectral resolution of data. Some model examples, for the spectra of NGs, S2Gs and SFGs, are given in Fig.~2. After masking the wavelength ranges affected by strong emission lines, we modeled the spectra with {\small STARLIGHT}, thus estimating the weight of each SSP contribution to the integrated signal and reproducing pure stellar spectra.

Applying this analysis to the whole sample, we observe that, on average, the properties of stellar populations differ significantly in the nuclei of distinct galaxy classes. Fig.~3 illustrates this general result in a plot of the average spectra of SFGs, S2Gs, and NGs, for comparison, after normalization to the specific flux at 5500~\AA. According to expectations, the spectra of SFGs are clearly different, with respect to the other classes, due to the presence of young hot stars, which dominate the emission, yielding a blue peaked spectral energy distribution (SED). The spectra of NGs and S2Gs, on the other hand, are remarkably similar and characteristic of an evolved stellar population. In spite of the comparable SEDs, however, S2Gs show on average a 4000~\AA\ break that is clearly shallower than the corresponding feature of NGs. It becomes therefore evident that an excess of hot stars, specifically belonging to the spectral class A, populates the circum-nuclear region of active galaxies.
\begin{figure}
\begin{center}
\includegraphics[width=8.4cm]{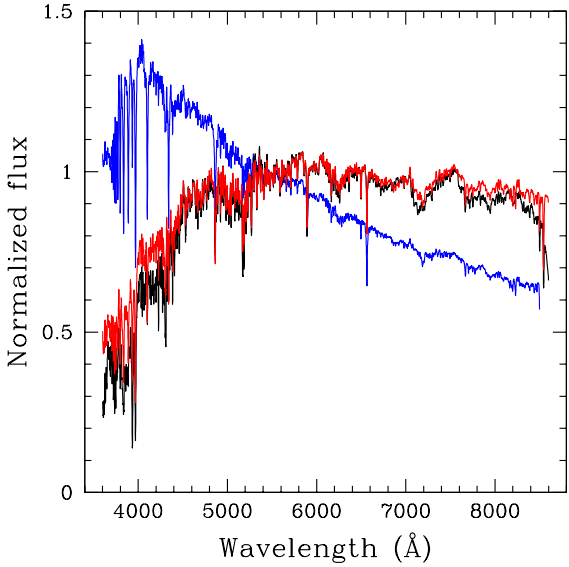}
\end{center}
\caption{Average normalized spectra of the stellar component in the circum-nuclear regions of NGs (black line), S2Gs (red line) and SFGs (blue line).\label{f02}}
\end{figure}

In the following sections we shall discuss how the average properties detected in the sample affect the central environment of the objects under investigation. The focus of our analysis will be centered on the region covered by the $2.4 - 4.8\,$kpc aperture of the SDSS fiber including the galaxy nuclei. We dub such region the circum-nuclear part of the galaxy, while we talk about {\it nuclear processes} to refer to the unresolved source of ionizing radiation in the nucleus.

\subsection{Gas components}
Once the stellar contributions in our sample of spectra had been reproduced by means of proper templates, we turned our attention to the analysis of the diffuse ISM. This task requires different approaches for the various classes of galactic nuclei we previously distinguished. Indeed, the physical processes, which are relevant to ionized gas in presence of either an AGN or young stars, represent far extremes of the ionization phenomenology and they involve quite different gas phases. Each class must, therefore, be dealt with by means of the most appropriate technique, though tests performed by comparing the results of multiple methods on the same objects have also been taken into account.

\subsubsection{Photoionization models}
Modeling the physics of ionized gas emitting regions in AGNs is a great challenge. Following \citet{Netzer08}, there are five main aspects to be taken into account for a correct representation of the ionized gas: photoionization and radiative recombination, thermal balance, ionizing spectrum, gas chemical composition, and cloud or filament distribution. In addition, the thermal balance has to consider mechanical heating and the role of dust, which is always present in the narrow line regions (NLR) of AGNs. We have excluded from our analysis gas kinematics, winds, and gas confinement mechanisms. Exploiting photoionization codes, such as {\small CLOUDY} v06.02 \citep[see][]{Ferland98}, we are able to solve numerically the ionization and thermal structure of a single cloud. However, we can also explore different parameter spaces and build more sophisticated models, such as composite models with more clouds and various geometries, as well as time dependent models. In our calculations we assumed open geometry. The input parameters, that we are considering, include the ionizing spectrum (a non-thermal power law, in our case), the density $n_e$, the ionized column density $N_c$(\ion{H}{II}), the ionization parameter $U$, and the metal abundances $Z / Z_\odot$. The ionization parameter expresses the ionization degree of photo-ionized gas and it is defined as the ratio of the ionizing photon density and the gas density, according to the relation:
$$U = \frac{Q({\rm H})}{4\pi r^2 c n_{\rm H}} \eqno(3)$$

\begin{table}
\begin{center}
\caption{Single cloud model parameter space}
\begin{tabular}{cccccc}
\hline
\hline
$Z / Z_\odot$ & $D / G$ & $\alpha$ & $\log(U)$ & $\log(n_e)$ & $\log[N_c$(\ion{H}{II})] \\
\hline
0.5 & 0.25 & -1.9 & -3.6 & 1.0 & 19.0 \\ 
1.0 & 0.50 & -1.6 & -3.2 & 1.5 & 19.4 \\
1.5 & 0.75 & -1.3 & -2.8 & 2.0 & 19.8 \\
2.0 & 1.00 & -1.0 & -2.4 & 2.5 & 20.2 \\
2.5 &      &      & -2.0 & 3.0 & 20.6 \\
3.0 &      &      & -1.6 & 3.5 & 21.0 \\
3.5 &      &      &      & 4.0 & 21.4 \\
    &      &      &      &     & 21.8 \\
    &      &      &      &     & 22.2 \\
\hline
\hline
\end{tabular}
\end{center}
\end{table}
In this work we exploited three kinds of photoionization models: a set of single cloud models and two of composite models, based on a combination of different single cloud models. The calculations were developed to explore the parameter space, according to the procedure described in \citet{Vaona10}. Actually, in the literature there are two fundamental approaches regarding composite models, either considering a dense cloud inside a medium with low density \citep{Binette96}, or combining separately two clouds \citep{Komossa97}. Since our aim is to estimate the mean physical parameters, which characterize the NLRs, using spectra in the visible range, a particular attention is put on the metallicity, because this parameter is important in the investigation of the nature of gas and of its origin. We generated a great number of models, in order to fit all the observed lines, subsequently applying a $\chi^2$ test to compare our simulations with the observed fluxes.

In order to explore the 5-dimensional parameter space we built a set of single cloud models with the parameter values reported in Table 1. The ionizing spectra were assumed to be power laws with $F_\nu \propto \nu^\alpha$ in the range $10\, \mu$m - $50\,$keV, a cut-off at low energies ($\nu^{2.5}$) and another at high energies ($\nu^{-2}$). This choice is in agreement with the results given by \citet{Komossa97}, \citet*{Groves04} and supported by X-ray observations \citep[see][]{Mainieri07}. The basic model is conceived to evaluate the ionizing photon flux across a slab with a fixed density, column density and dust to gas ratio ($D / G$). The distance of the source does not need to be determined, since the ionization parameter is fixed. The intrinsic spectrum is calculated using CLOUDY and the calculation stops when either the fixed required column density has been reached, or the temperature in the simulation has fallen below 4000~K. We assume that the intrinsic spectrum is reddened by the ISM, so that, when the models are compared to the observed spectra, a correction for extinction must be taken into account.

When the ionizing spectrum is a non-thermal power law, the ionization structure is very complex. The partially ionized region is so large that different ionization states are mixed and the solution of the ionic abundances, which make up the total abundances, gets quite complicated. An appropriate choice of the chemical abundance sets is a critical step in the creation of the models, because they are strongly connected with another fundamental aspect of the simulations, specifically the role of dust and its composition. Once a reference sample of different abundances is assumed (usually the solar set of values), we need a law describing the change in the chemical composition that will be used for the exploration of different metallicities. Observations of the ionized gas chemical abundances in \ion{H}{II} regions and starburst galaxies show that all elements, except Nitrogen and Helium, in first approximation are found in fixed proportions \citep*[e. g.][and references therein]{Masegosa94, VanZee98, Peimbert10}. Usually, the \ion{H}{II} metallicities are expressed in terms of Oxygen abundance, because in the visible spectra it is possible to measure three different ionization states of Oxygen, which is the most abundant element after Hydrogen and Helium. Then, we can assume that all elements scale directly with O abundance, except for N and He that must be treated separately. The abundance of N, as well as He, must be corrected for secondary production \citep[see e. g.][]{VilaCostas93}:
$$\frac{X({\rm N})}{X({\rm H})} = \frac{X({\rm O})}{X({\rm H})} [10^{-1.6} + 10^{2.37 + \log({\rm O}/{\rm H})}] \eqno(4a)$$
$$\frac{X({\rm He})}{X({\rm H})} = 0.0737 + 0.0293 \frac{Z}{ Z_\odot}. \eqno(4b)$$

The composition and quantity of dust are also sources of uncertainties. The set of abundances needs to take into account both gas and dust composition in order to maintain the total assumed element abundances. As for the abundances of the elements, whose reference values are assumed to be similar to the solar values, the default values of abundance and composition for dust are taken from the Galactic ISM. Chosen the features of the dust, the abundances of the gas component are calculated by subtracting the dust abundances from the total. With the exception of the noble gases, all abundances must be corrected for depletion into dust. The determination of these corrections is a main issue of modeling, since the only possible assumptions are based on the observational properties of the Galactic ISM. Another important assumption involves $D / G$, which we again express in terms of its value relative to the local ISM. By changing the relative internal dust content as a function of the metallicity, the depletion factors must also be changed. A method to estimate the depletion factors with the variation of the dust content is presented in \citet{Binette93}. It has been suggested that the depletion into grains could be constant in many directions of the Galaxy \citep[see][]{Vladilo02}, so we adopted the depletion factors reported in \citet*{Groves06}, as reference values, and a linear law for scaling the dust abundance with metallicity. Using the conventional logarithmic notation, the relation between depletion factor and gas abundance is given by:
$$depl(X) = \log (X / {\rm H})_{gas} - \log (X / {\rm H})_{tot}. \eqno(5)$$

\begin{table}
\begin{center}
\caption{Abundances and depletion factors for $D / G = 1.00$}
\begin{tabular}{ccccc}
\hline
\hline
$Z / Z_\odot$ & O / H & (N / H)$_{tot}$ & (N / H)$_{gas}$ & {\it depl} \\
\hline
0.5 & -3.64 & -4.75 & -5.54 & -0.79 \\
1.0 & -3.34 & -4.22 & -4.52 & -0.30 \\
1.5 & -3.16 & -3.89 & -4.09 & -0.19 \\
2.0 & -3.04 & -3.66 & -3.80 & -0.14 \\
2.5 & -2.94 & -3.48 & -3.59 & -0.11 \\
3.0 & -2.86 & -3.32 & -3.41 & -0.09 \\
3.5 & -2.80 & -3.19 & -3.27 & -0.08 \\
\hline
\hline
\end{tabular}
\end{center}
\end{table}
In this work, we assume $\log (X / {\rm O}) = const.$ for all chemical elements, with the exception of N and He. For these elements, we follow indications by \citet{Groves04}, employing a linear combination of the primary and secondary components of N with the requirement to match the adopted solar abundance patterns \citep*{Asplund05}. We focused our attention on the high metallicity models, because nuclear gas and Seyfert galaxies very rarely present sub-solar metallicities \citep{Groves06}. For each adopted $D / G$ we changed the depletion factor set in order to retain the total abundances. The N depletion factor must change with metallicity because, in agreement with our assumptions, if the dust increases with metallicity, there must be an excess of N in gas form due to Eq.~(4a). In Table~2 we report the depletion factors used for models with local ISM dust content ($D / G = 1$). If this assumption is correct, a great increment of N gas abundance for $Z / Z_\odot > 1.5$ should be observed, suggesting that it is not necessary to invoke an excess of N in order to match the observed [\ion{N}{II}]~$\lambda6584$ line intensity. It is important to stress that the depletion of the elements must be taken into consideration even if the nebula is without dust, such as in the radiation-pressure dominated models of \citet{Groves04}, according to which the dust is blown away by the wind. In our models, we assumed only two types of dust: graphite and silicates; their reference abundances are [C/H] = $1.22 \cdot 10^{-4}$ when $D / G = 1$, in the graphite, and [O/H] = $1.94 \cdot 10^{-4}$, [Mg/H] = $3.15 \cdot 10^{-5}$, [Si/H] = $2.82 \cdot 10^{-5}$, [Fe/H] = $2.70 \cdot 10^{-5}$ for silicate, respectively, while the distribution of grain size is in the range [0.005, 0.25] $\mu$m.

The single cloud model is only the first step of our modeling effort, since in the observed Seyfert spectra two distinct components are usually detected, yielding high and low ionization lines. More accurate simulations can be produced introducing a double component model with two clouds, each accounting for either the high or the low ionization lines. We considered all the possible combinations between two clouds with fixed power law indices and metallicities. The spectral lines emitted by the first and second cloud are combined by a weighted mean, where the weight is given by the product of H$\beta$ luminosity ratio and the ratio of the solid angles subtended by the clouds, as seen from the source of ionizing radiation ($\omega_1 / \omega_2$, called geometrical factor and henceforth indicated with $GF$). The combined fluxes were derived from the geometry of the system. Given a line luminosity:
$$L = 4 \pi r^2 S \omega / 4\pi  \eqno(6)$$
where $r$ is the radius of the nebula, $\omega / 4 \pi$ is the covering factor, $S$ the emission line intensity (${\rm erg\, s^{-1}\, cm^{-2}}$), from the relation between flux and luminosity, we get:
$$4 \pi d^2 F = 4 \pi r^2 S \omega / 4 \pi  \eqno(7)$$
where $F$ is the observed flux (${\rm erg\, s^{-1}\, cm^{-2}}$) and $d$ is our distance to the system. Considering two clouds, from the definition of ionization parameter in Eq.~(3), we have that:
$$\frac{U_1}{U_2} = \frac{Q({\rm H_0})}{4 \pi r_1^2 n_1 c} \frac{4 \pi r_2^2 n_2 c}{Q({\rm H_0})} \eqno(8)$$
so that:
$$\frac{r_2^2}{r_1^2} = \frac{U_1 n_1}{U_2 n_2}. \eqno(9)$$
From Eq.~(7) we also have the flux expressed as:
$$F_i = \frac{r_i^2 S_i \omega_i}{4 \pi d^2} \eqno(10)$$
and
$$\frac{F_2}{F_1} = \frac{r_2^2}{r_1^2} \frac{S_2 \omega_2}{S_1 \omega_1}, \eqno(11)$$
which, recalling Eq.~(9), finally gives:
$$\frac{F_2}{F_1} = \frac{U_1 n_1 S_2 \omega_2}{U_2 n_2 S_1 \omega_1}. \eqno(12)$$
In terms of H$\beta$, this ratio can be written as:
$$\frac{F_2({\rm H}\beta)}{F_1({\rm H}\beta)} = GF \cdot \frac{L_2^*({\rm H}\beta)}{L_1^*({\rm H}\beta)}, \eqno(13)$$
where $L_2^*({\rm H}\beta) / L_1^*({\rm H}\beta)$ is the H$\beta$ luminosity ratio if $\omega_1 = \omega_2$. These quantities are calculated by CLOUDY for each model. If we introduce the emission line intensity $I_\lambda$ relative to H$\beta$:
$$I_i(\lambda) = \frac{F_i(\lambda)}{F_{tot}({\rm H}\beta)}, \eqno(14)$$
where $F_{tot}({\rm H}\beta)$ is the total observed flux of H$\beta$, we obtain:
$$I_{tot}(\lambda) = \frac{I_2(\lambda) + I_1(\lambda) F_1({\rm H}\beta) / F_2({\rm H}\beta)}{1 + F_1({\rm H}\beta) / F_2({\rm H}\beta)} \eqno(15)$$
and finally:
$$I_{tot}(\lambda) = \frac{I_2(\lambda) + I_1(\lambda) GF \cdot L_1^*({\rm H}\beta) / L_2^*({\rm H}\beta)}{1 + GF \cdot L_1^*({\rm H}\beta) / L_2^*({\rm H}\beta)} \eqno(16)$$
If $GF$ goes to zero, only the second cloud is visible; on the contrary, if $GF$ goes to infinity, only the first cloud is visible. Five values of $GF$ have been used (0.25, 0.5, 1, 2 and 4), so that each pair of models provides five new synthetic spectra. The total number of models obtained with the constraints on $D / G$, $\alpha$, $Z / Z_\odot$, fixed for each pair of clouds, is approximately $9 \cdot 10^6$.

\subsubsection{The P-index method}
The determination of chemical abundances in the \ion{H}{II} regions of SFGs is complicated by the difficulty of getting reliable measurements of the temperature of the \ion{H}{II} clouds. The auroral lines, such as [\ion{O}{III}] ~$\lambda$4363, [\ion{S}{III}]~$\lambda$6312 and [\ion{N}{II}]~$\lambda$5755, used to evaluate the temperature, are usually weak, especially in low-excitation metal-rich \ion{H}{II} regions. For this reason, the direct method is almost useless in the vast majority of cases. \citet{Pagel79} suggested that some empirical calibrations between the nebular line fluxes could be used to derive directly the O abundance or the temperature of \ion{H}{II} regions. Specifically, the parameter defined as:
$$R_{23} = \frac{F([{\rm O II}] \lambda 3727) + F([{\rm O III}] \lambda\lambda 4959,5007)}{F({\rm H}\beta)} \eqno(17)$$
was used to measure directly the abundance of O. This ratio was re-calibrated using photoionization models and confirmed through other observations. Nonetheless, the comparison between the results obtained from different samples of galaxies remains very difficult because of the high heterogeneity of the model parameters and data. In all cases, the one-dimensional calibrations are systematically wrong as shown by \citet{Pilyugin05}. The $R_{23}$ index alone does not remove all the degeneracies, so a more general two-dimensional parametric calibration, called the P-index, was proposed, introducing a new excitation parameter:
$$P_{23} = \frac{F([{\rm O III}] \lambda\lambda 4959,5007)}{F([{\rm O II}] \lambda 3727) + F([{\rm O III}] \lambda\lambda 4959,5007)}. \eqno(18)$$

\citet{Pilyugin05} used their collection of high precision measurements to revise the P-index calibration, suggesting that the O abundance of \ion{H}{II} regions in SFGs could be directly estimated, without degeneracies, from:
$$12 + \log ({\rm O / H}) =$$
$$= \frac{(R_{23} + 726.1 + 842.2 \cdot P + 337.5 \cdot P^2)}{(85.96 + 82.76 \cdot P + 43.98 \cdot P^2 + 1.793 \cdot R_{23})} \eqno(19)$$

\subsection{Stellar masses}
The correlations between the Super Massive Black Hole (SMBH) masses and the properties of the host stellar bulge, such as its luminosity, mass, and velocity dispersion, represent the strongest argument in support for a common evolutionary track. Investigating these relationships in detail, however, requires large samples of high precision observations, in order to properly calibrate empirical laws on the observed data. Combining the spectroscopic and photometric information, collected in the SDSS database, \citet{Kauffmann03a} developed an useful technique to estimate the stellar masses of AGN host galaxies. They extensively exploited stellar population synthesis, in order to reproduce the observed colors and luminosities, thus deriving the most appropriate mass-luminosity ratios for these objects. We adopted the updated results of \citet{Salim07} as the stellar masses of our S2G and SFG samples.

\begin{figure}
\begin{center}
\includegraphics[width=8.4cm]{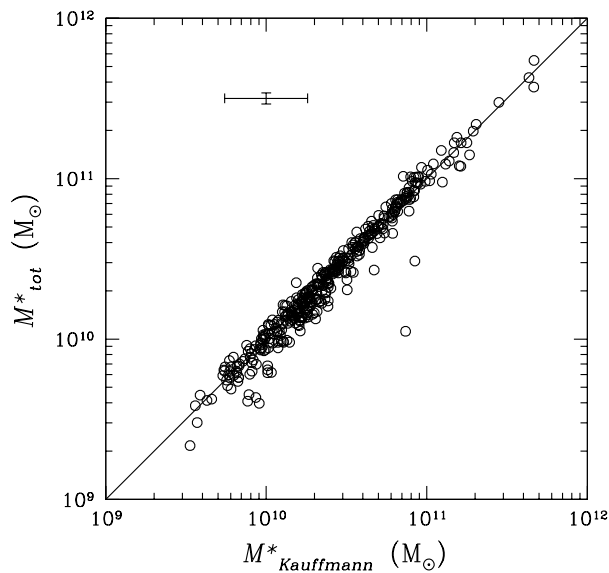}
\caption{Comparison between the average mass estimates derived by \citet{Salim07} and the stellar masses computed by means of Eq.~(20). The continuous line represents the perfect identity relationship, while the error bars describe the median uncertainty estimates derived from the spectral synthesis technique of \citet{Salim07} and that descending from the photometric measurement errors.}
\end{center}
\end{figure}
The case of normal galaxies, however, is somewhat different, since most of these objects are not included in any similar catalog, due to the absence of prominent emission lines in their nuclei. To estimate the stellar mass range covered by this fraction of our sample, we started from the results of our stellar population synthesis, which found no evidence of peculiarities in the stellar populations located in the circum-nuclear regions of NGs. From the stellar population synthesis of the spectra collected by the SDSS instrumentation, we derived the stellar masses in the circum-nuclear regions of these galaxies and their characteristic $M / L$ \citep{CidFernandes05}. Then, assuming that $M / L$ is constant throughout the galaxy, we computed the stellar masses from:
$$\frac{M^*_{tot}}{M^*_{fiber}} = 10^{0.414 (r'_{fiber} - r'_{tot})}, \eqno(20)$$
where $M^*_{tot}$ and $M^*_{fiber}$ are, respectively, the total mass in stars and the estimated stellar mass in the area covered by the fiber, while $r'_{tot}$ and $r'_{fiber}$ represent the $r'$ band Petrosian magnitude and the magnitude within the fiber. In principle, the uncertainty to be associated with this mass estimate results from two factors: at first, we have the photometric errors in the measurement of the fiber magnitude and in the assumption of the model representing the total galaxy flux; on the other hand there is the assumption of a constant representative $M / L$ ratio across the whole galaxy. In order to test the reliability of our method, we compare the mass estimates, derived with the photometric uncertainty only, with the results of \citet{Salim07}, for the objects included in both samples (Fig.~4). Given the small uncertainty range of the photometric measurements, we expect that the the scatter in the correlation between the two methods is mostly a consequence of the assumption on $M / L$. The comparison of results gives a high degree of correlation ($R = 0.978$), which supports the use of the photometric measurements to provide an estimate of the total stellar masses.

\begin{table}
\begin{center}
\caption{List of the emission lines calculated in the models}
\begin{tabular}{p{0.9cm} p{0.9cm} p{0.9cm} p{0.9cm} p{0.9cm} p{0.9cm}}
\hline\hline
Ion & $\lambda$ (\AA) & Ion & $\lambda$ (\AA) & Ion & $\lambda$ (\AA) \\
\hline
[\ion{O}{II}] & 3727 & [\ion{Ar}{IV}] & 4740 & [\ion{O}{I}] & 6363 \\
\end{tabular}
\begin{tabular}{p{0.9cm} p{0.9cm} p{0.9cm} p{0.9cm} p{0.9cm} p{0.9cm}}
[\ion{O}{II}]   & 3729 & H$\beta$        & 4861 & [\ion{N}{II}]   & 6548 \\
\end{tabular}
\begin{tabular}{p{0.9cm} p{0.9cm} p{0.9cm} p{0.9cm} p{0.9cm} p{0.9cm}}
[\ion{Ne}{III}] & 3869 & [\ion{O}{III}]  & 4959 & H$\alpha$       & 6563 \\
\end{tabular}
\begin{tabular}{p{0.9cm} p{0.9cm} p{0.9cm} p{0.9cm} p{0.9cm} p{0.9cm}}
[\ion{Ne}{III}] & 3968 & [\ion{O}{III}]  & 5007 & [\ion{N}{II}]   & 6584 \\
\end{tabular}
\begin{tabular}{p{0.9cm} p{0.9cm} p{0.9cm} p{0.9cm} p{0.9cm} p{0.9cm}}
[\ion{S}{II}]   & 4064 & [\ion{N}{I}]    & 5200 & [\ion{S}{II}]   & 6716 \\
\end{tabular}
\begin{tabular}{p{0.9cm} p{0.9cm} p{0.9cm} p{0.9cm} p{0.9cm} p{0.9cm}}
H$\gamma$       & 4340 & [\ion{Fe}{VII}] & 5721 & [\ion{S}{II}]   & 6731 \\
\end{tabular}
\begin{tabular}{p{0.9cm} p{0.9cm} p{0.9cm} p{0.9cm} p{0.9cm} p{0.9cm}}
[\ion{O}{III}]  & 4363 & \ion{He}{I}     & 5876 & [\ion{Ar}{III}] & 7135 \\
\end{tabular}
\begin{tabular}{p{0.9cm} p{0.9cm} p{0.9cm} p{0.9cm} p{0.9cm} p{0.9cm}}
\ion{He}{II}    & 4686 & [\ion{Fe}{VII}] & 6087 & [\ion{O}{II}]   & 7325 \\
\end{tabular}
\begin{tabular}{p{0.9cm} p{0.9cm} p{0.9cm} p{0.9cm} p{0.9cm} p{0.9cm}}
[\ion{Ar}{IV}]  & 4711 & [\ion{O}{I}]    & 6300 &                 & \\
\hline
\hline
\end{tabular}
\end{center}
\end{table}
\section{Results and discussion}
\subsection{Comparison of models and observations}
To study the gas phase metallicities of active galaxies, we compared the observed emission lines with the synthetic spectra, produced by our models. In Table~3 we give a list of the lines that we included in the model calculations. In general we did not require that all the listed emission lines could actually be detected in the spectra, since some of them could be intrinsically weak or absent. In order to apply different models to specific spectra, we only required that the number of detected emission lines was high enough to constrain the parameter space of the models in use. We spent a considerable effort to reproduce as effectively as possible the observed spectra, with the aim of determining the most realistic set of input parameters. \citet*{Oliva99} proposed a new method for deriving abundances in the NLR of active galaxies, consisting in a selection of fair models from a large set (27000 models, in their work), which is capable of fitting the observed line spectra, and then slightly modifying the chemical abundances in order to determine the best estimates of the metallicity of the AGN. Our approach is slightly different. Instead of varying the input parameters, we produced a very large number of synthetic spectra, which we employed to determine the set of models, yielding the most accurate synthetic spectral features for each observed spectrum, with a $\chi^2$ test. In this case, the  $\chi^2$ is given by:
$$\chi^2 = \sum_i \frac{(F_\lambda^{obs} - F_\lambda^{mod})^2}{2 \sigma_i^2}, \eqno(21)$$
where $F_i^{obs} - F_i^{mod}$ is the difference between observed and modeled flux of the $i$-th line and $\sigma_i$ the associated error. Therefore, the most probable set of input parameters can be determined for each observed spectrum. The comparison has been carried out taking into account only the synthetic lines with a predicted intensity corresponding to a $S / N > 3$ detection in the observed spectrum. Lines below the $S / N$ observational threshold are excluded from the residual calculation. Another important concern, in the comparison of models with observed spectra, is the problem of intrinsic extinction. While the observed emission lines are affected by a certain amount of reddening, the model spectra are extinction free. Therefore, before performing the $\chi^2$ test, we artificially reddened our model spectra in order to achieve the observed $F({\rm H\alpha}) / F({\rm H\beta})$ ratio. To do this, we combined $F({\rm H\alpha}) \pm \Delta F({\rm H\alpha})$ with $F({\rm H\beta}) \pm \Delta F({\rm H\beta})$ and we evaluated 9 different Balmer line ratios on the observed spectra. Assuming a wavelength dependent extinction law in the form related to the $A_V$ parameter \citep*{Cardelli89}, we derived all the extinction curves giving rise to a non-negative value of $A_V$. In this way, we produced up to 9 different synthetic spectra for each model.

The synthetic spectra are expressed in a relative flux scale, normalized to the predicted flux of H$\beta$, $F({\rm H}\beta)$, which acts as a scale factor. To compare these data with observations, we converted the synthetic spectra in an absolute flux scale, by means of this factor. The accepted model has to satisfy a predetermined significance level, suggested by the analysis of the reduced $\chi^2$ distribution, given by:
$$\chi_{red}^2 = \frac{\chi^2}{n_r - n_p}, \eqno(22)$$
where $n_r$ is the number of measured lines and $n_p$ the number of parameters used in the models. In order to consider the models a good approximation of the observed spectra, the $\chi_{red}^2$ distribution of the residuals left by the best fit models over the sample of observations must tend to a Gaussian function, peaked at 1. A peak located at much smaller values would probably indicate an over-fitting, where non significant parameter sets fit very large uncertainty ranges. On the other hand, if the peak falls at larger values, the models cannot reproduce the observed spectra, or the errors are too small. After several tests, we found that the best match between models and observations could be carried out assuming the observed flux uncertainties in Eq.~(21) at the 2$\sigma$ level. Since the model parameters affect each other in a complex interplay, we had to estimate the relative uncertainties by introducing a confidence range of 10 per cent around the best fit values. All the models falling in the resulting parameter space region were taken into account and those meeting the condition:
$$\chi^2 - \chi_{min}^2 < \frac{\chi^2_{min}}{4} \eqno(23)$$
\citep[corresponding to a 25 per cent level of confidence, but see][]{Molla02}
were considered acceptable as well. The best fit parameters and their uncertainties are evaluated by computing the average and r.m.s. values from all the acceptable models. 

\begin{figure}
\begin{center}
\includegraphics[width=8.4cm]{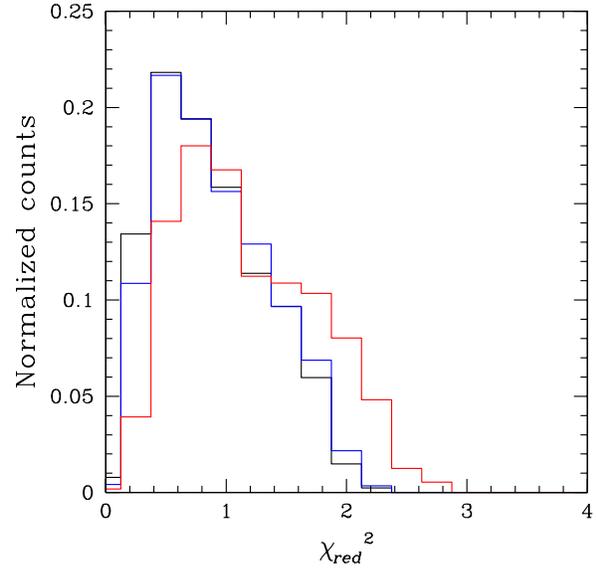}
\caption{The $\chi_{red}^2$ distribution for the $D / G = 1$ best fit models carried out by single cloud (blue histogram), two clouds (red histogram), and Binette's models (black histogram) over the S2G galaxy sample.}
\end{center}
\end{figure}
In our analysis, we consider 7 free parameters for the single cloud models and 11 for the two-cloud ones. As a consequence, more than 7 measured lines are required to fit the single cloud models, while at least 12 measured lines must be available for the composite models. About 50\% of the spectra shows less than 12 measured lines, meaning that the two-cloud models can be applied only to 50\% of the original galaxy sample. This fraction, however, includes spectra where $S / N$ is high enough to detect also several weak lines above the 3-$\sigma$ significance level. Single cloud models, on the other hand, have difficulties to reproduce with reasonable accuracy spectra with many observed emission lines. Since the simple rejection of spectra that we are not able to fit with some models would raise a strong bias on the inferred properties of our sample, we studied the different model efficiencies in fitting the observed spectra. To achieve this, we split the S2G sample in two groups, namely spectra with $n_r \geq 12$ and spectra with $n_r < 12$. Since the $D / G$ ratio is not considered as a model parameter, the models with different $D / G$ values are treated separately, so that we are left with four families of models (namely $D / G$ = 0.25, 0.50. 0.75, and 1 times the Galactic reference). On a total of 2153, 1141 spectra have $n_r < 12$ and the remaining $n_r \geq 12$. For this reason, the comparison between single cloud and two-cloud models is only possible for 1012 spectra. The total percentage of spectra, which produced satisfactory fits, are $\sim 65$ per cent and $\sim 50$ per cent with the single and two-cloud models respectively. The percentage of spectra successfully fitted with $n_r < 12$ is 80 per cent, while $\sim 45$ per cent is the fraction of spectra fitted with $n_r \geq 12$. The $\chi_{red}^2$ distributions for the $D / G = 1$ model families are shown in Fig~5. The distributions of the single cloud and Binette's models for high $D / G$ peak at $\chi_{red}^2 \approx 0.6$, while the two-cloud models show tails of high residuals, although their peak is closer to the desired value of $\chi_{red}^2 = 1$. 

\begin{figure}
\begin{center}
\includegraphics[width=8.4cm]{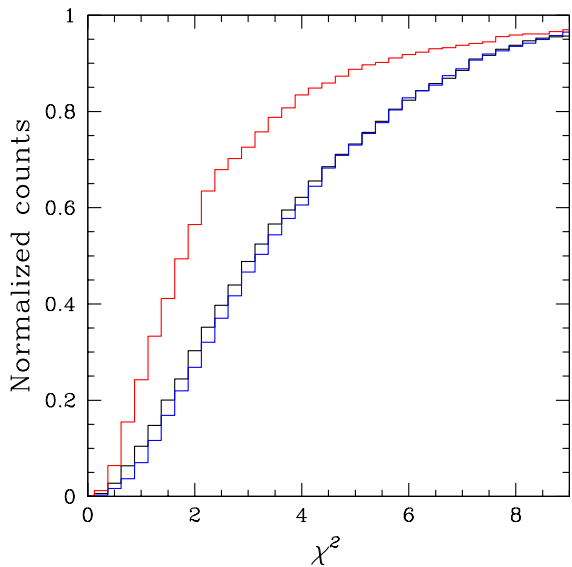}
\caption{The cumulative $\chi^2$ distribution for the $D / G = 1$ families of single cloud (blue histogram), two cloud (red histogram), and Binette's models (black histogram) applied to the shared spectra.}
\end{center}
\end{figure}
A more qualitative difference between the model families becomes evident by comparing the distribution of absolute $\chi^2$ illustrated in Fig.~6, since this statistics represents the goodness of the overall fitting procedure (the input model parameters, the choice of the significance level and the errors) and it is possible to discriminate between two models able to fit the same spectrum. For each assumed $D / G$ value, there are respectively 289, 356, 410, and 396 spectra, which were fitted by single cloud, two-cloud and Binette's models, but with dramatically different $\chi^2$ distributions. The average values of $\chi_{red}^2$ for each group, with the corresponding standard deviations, are listed in Table~4, but the differences can be better appreciated looking at the cumulative distributions of $\chi^2$ plotted in Fig.~6: when applied to the same spectra, two-cloud models produce lower residuals than single cloud or Binette's models. This is reasonable because spectra with a larger number of measured lines could be more effectively fitted with an increased parameter resolution, particularly for the power law index and the ionization structure, at the cost of a higher computational load. Anyway, these considerations should be cautiously handled, since the percentage of successful fits are calculated averaging over few spectra, so that a single measurement could drastically modify the result. The most affected lines are those from [\ion{Ar}{IV}] $\lambda\lambda4711, 4740$, [\ion{N}{I}] $\lambda5200$, [\ion{Fe}{VII}] $\lambda\lambda5721, 6087$, \ion{He}{I} $\lambda5876$, and [\ion{Ar}{III}] $\lambda7135$.

\subsection{The properties of gas}
In Table 5 we summarize all the mean parameter values giving the best fit solutions, with their 1$\sigma$ uncertainty ranges, arranged according to the different model families. We observe that the power-law index $\alpha$ of the ionizing spectrum decreases to -1.8, as soon as we increase the $D / G$ ratio, probably because of the decreasing gas metallicity content (a large fraction of metals is depleted into grains). Indeed, a low metallicity of the gas implies a much slower cooling and a weaker ionizing spectrum can produce the same degree of ionization. The total metallicity is well defined in all models with negligible scatter: $Z = 1.5 - 2.5\, Z_\odot$ is the range of fitted values with the single cloud models, while the two-cloud models point towards estimates of $Z = 1.0 - 2.0\, Z_\odot$. The high metallicity of the gas indicates that the AGN fuel is an evolved material. This, in turn, supports the idea that the gas has a local origin in the vast majority of the observed S2Gs. The densities are in agreement with the expected values, since single cloud and two-cloud models produce similar density estimates, both consistent with the indications obtained from the [\ion{S}{II}] $\lambda\lambda 6716, 6731$ intensity ratios. The column densities are very similar in the single cloud models and in the first cloud of the two-cloud models, while the second cloud of the two-cloud models has a lower column density and a lower ionization parameter, as well, allowing us to treat such cloud as an ionization bounded case. The ionization parameter in the single cloud models is almost constantly fixed at $\log (U) = -3$ for different values of $D / G$. In two-cloud models the high ionization cloud shows a little variation of the ionization parameter, depending on $D / G$, with a mean value of $\log (U) = -2.6$, while very steady values of $\log (U) = -3.3$ are found for the low ionization cloud. In general, the amount of ionization in the single cloud models is intermediate between the values found in the first and second cloud of the two-cloud models.
\begin{table}
\begin{center}
\caption{Comparison of the average residuals left by single and two-cloud models on the shared spectra.}
\begin{tabular}{cccc}
\hline\hline
$D / G$ & $N_{\rm spectra}$ & $\langle \chi_{red}^2 \rangle_{\rm single\, cloud}$ & $\langle \chi_{red}^2 \rangle_{\rm two-cloud}$ \\
\hline
0.25 & 289 & 0.929 $\pm$ 0.373 & 1.207 $\pm$ 0.590 \\
0.50 & 356 & 0.914 $\pm$ 0.398 & 1.182 $\pm$ 0.636 \\
0.75 & 410 & 0.922 $\pm$ 0.418 & 1.140 $\pm$ 0.607 \\
1.00 & 396 & 0.862 $\pm$ 0.407 & 1.112 $\pm$ 0.578 \\
\hline
\end{tabular}
\end{center}
\end{table}

\begin{table*}
\begin{minipage}{160mm}
\begin{center}
\caption{Average output parameters and standard deviations for the four $D / G$ families of single cloud, two-cloud and Binette's photoionization models.}
\begin{tabular}{cccccccccc}
\hline\hline
Parameter & SC$_{0.25}$ & SC$_{0.50}$ & SC$_{0.75}$ & SC$_{1.00}$ & 2C$_{0.25}$ & 2C$_{0.50}$ & 2C$_{0.75}$ & 2C$_{1.00}$ & Binette \\
\hline
$\alpha$             & -1.5 $\pm$ 0.3 & -1.6 $\pm$ 0.3 & -1.7 $\pm$ 0.3 & -1.8 $\pm$ 0.3 & -1.3  $\pm$ 0.3 & -1.4 $\pm$ 0.3 & -1.6 $\pm$ 0.3 & -1.8 $\pm$ 0.3 & -2.0 $\pm$ 0.2 \\
$Z / Z_\odot$         & 2.0 $\pm$ 0.6 & 1.9 $\pm$ 0.6 & 1.9 $\pm$ 0.5 & 1.8 $\pm$ 0.5 & 1.6  $\pm$ 0.5 & 1.7 $\pm$ 0.4 & 1.7 $\pm$ 0.4 & 1.7 $\pm$ 0.4 & 1.5 $\pm$ 0.4 \\
$\log n_{e, 1}$       & 2.3 $\pm$ 0.7 & 2.3 $\pm$ 0.7 & 2.2 $\pm$ 0.7 & 2.1 $\pm$ 0.7 & 2.2  $\pm$ 0.8 & 2.2 $\pm$ 0.8 & 2.1 $\pm$ 0.7 & 2.2 $\pm$ 0.8 & n.a. \\
$\log N_{c, 1}$(H II) & 20.5 $\pm$ 0.7 & 20.3 $\pm$ 0.6 & 20.0 $\pm$ 0.5 & 19.9 $\pm$ 0.4 & 20.2 $\pm$ 0.4 & 20.2 $\pm$ 0.4 & 20.2 $\pm$ 0.4 & 20.2 $\pm$ 0.4 & 20.0 $\pm$ 0.4 \\
$\log U_1$           & -3.1 $\pm$ 0.2 & -3.1 $\pm$ 0.2 & -3.0 $\pm$ 0.2 & -3.0 $\pm$ 0.3 & -2.7  $\pm$ 0.4 & -2.6 $\pm$ 0.4 & -2.6 $\pm$ 0.4 & -2.5 $\pm$ 0.4 & -1.6 $\pm$ 0.2 \\
$\log n_{e, 2}$       & n.a. & n.a. & n.a. & n.a. & 3.0  $\pm$ 0.9 & 2.8 $\pm$ 0.9 & 2.7 $\pm$ 0.9 & 2.5 $\pm$ 0.8 & n.a. \\
$\log N_{c, 2}$(H II) & n.a. & n.a. & n.a. & n.a. & 19.7  $\pm$ 0.4 & 19.7 $\pm$ 0.3 & 19.7 $\pm$ 0.3 & 19.7 $\pm$ 0.3 & n.a. \\
$\log U_2$                & n.a. & n.a. & n.a. & n.a. & -3.3  $\pm$ 0.2 & -3.3 $\pm$ 0.2 & -3.3 $\pm$ 0.2 & -3.3 $\pm$ 0.3 & -3.5 $\pm$ 0.3\\
$GF$                 & n.a. & n.a. & n.a. & n.a. & n.a. & n.a. & n.a. & n.a. & 4.1 $\pm$ 2.1 \\
$A_V$                & 1.4 $\pm$ 0.6 & 1.5 $\pm$ 0.6 & 1.5 $\pm$ 0.6 & 1.6 $\pm$ 0.6 & 1.2 $\pm$ 0.4 & 1.2 $\pm$ 0.4 & 1.3 $\pm$ 0.4 & 1.4 $\pm$ 0.4 & 1.6 $\pm$ 0.6 \\
\hline\hline
\end{tabular}
\end{center}
\end{minipage}
\end{table*}
On the other hand, as we show in Fig.~7, a substantially different situation applies to the chemical properties of the ISM in star forming galaxies. According to the metallicity indicators discussed in \S3.2.2, these objects are rather characterized by ISM with a sub-solar chemical composition, suggesting that their on-going star formation is a first generation process. This result is also confirmed by the evidence that stellar masses in galaxies hosting nuclear star formation are generally lower than in S2Gs, in agreement with the findings of \citet{Kauffmann03b}.
\begin{figure}
\begin{center}
\includegraphics[width=8.4cm]{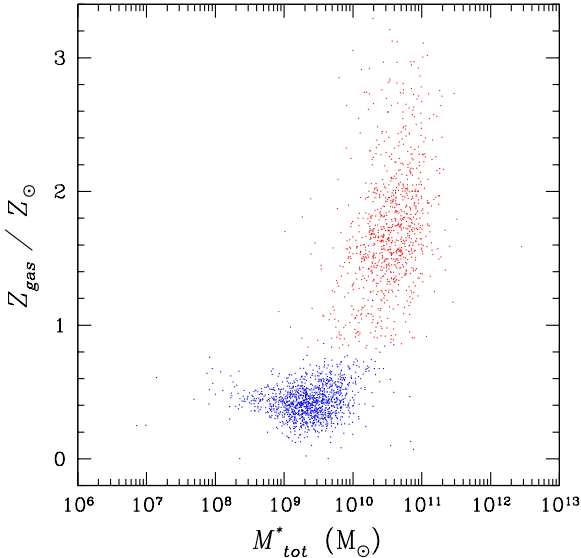}
\caption{Relationship between the estimated stellar masses and gas phase metallicities for SFGs (blue points) and S2Gs (red points).}
\end{center}
\end{figure}

\subsection{The properties of stars}
The processes of star formation and stellar feedback originate an interplay, which connects the chemical properties of stars with those of the ISM, wherein they reside. The inter-stellar gas is chemically enriched through stellar activity, while the metallicity of stars reflects the chemical composition of the environment, where they are formed. The spectral synthesis technique, which we exploited to account for the stellar continua and absorption features, reproduce the circum-nuclear stellar populations of our galaxy sample, in terms of a mixture of SSPs. Following \citet{CidFernandes05}, we estimate the average stellar population metallicities through:
$$\langle Z_\star \rangle = \sum_{i = 1}^{92} \mu_i Z_i, \eqno(24)$$
where $\mu_i$ is the mass fraction ascribed to the $i$-th SSP component of the spectrum. Considerations on the application of such paradigm to large samples led to evaluate a typical uncertainty of $\Delta \log Z_\star \sim 0.1$dex.

\begin{figure}
\begin{center}
\includegraphics[width=8.4cm]{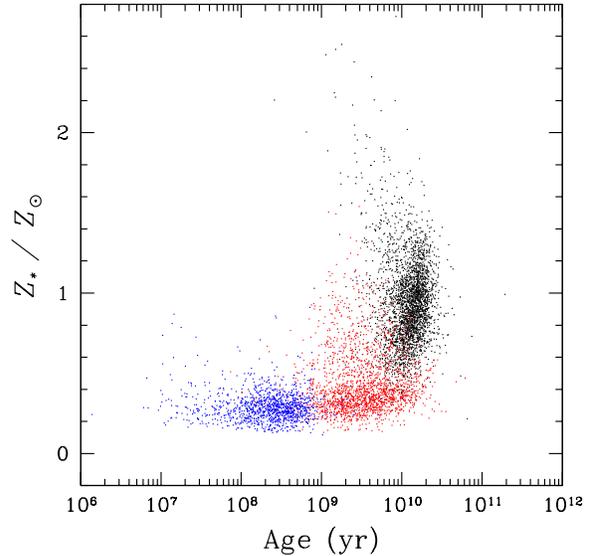}
\caption{Metallicity of the circum-nuclear stellar population, plotted as a function of the population age of SFGs (blue points), S2Gs (red points), and NGs (black points).}
\end{center}
\end{figure}
Despite the different chemical compositions of the gas seen in Fig.~7, the average metallicity detected in the stellar populations in SFGs and S2Gs is mostly similar, but with a slight over-abundance of heavy elements in the second group. As it is shown in Fig.~8, indeed, the metallicity of stars in the SFG sample stays low, with no appreciable dependence on the age of the populations, as we would expect for a generation of stars originated by an unpolluted medium. S2Gs, on the contrary, exhibit also a spread of high metallicity population components, which represents a hint of further star formation in a chemically enriched medium, bringing such objects on the same range of metallicities, commonly observed in the circum-nuclear regions of NGs. This result indicates that SFGs and S2Gs host stellar populations mainly formed in a pristine inter-stellar gas, but, as it is also suggested by the average spectral properties illustrated in Fig.~3, the process occurred earlier in S2Gs, where the O/B spectral class stars have already evolved off the main sequence. If this scenario is actually correct and we do observe stellar populations formed by pristine gas both in SFGs and S2Gs, then it is likely that the circum-nuclear regions of active galaxies experienced past star formation, on a time scale which should account for the evolution of massive and short-lived O/B stars and for the enrichment of the ISM, but not for that of lower mass A stars, as well as for the formation of new stars in the enriched medium.

\begin{figure}
\begin{center}
\includegraphics[width=8.4cm]{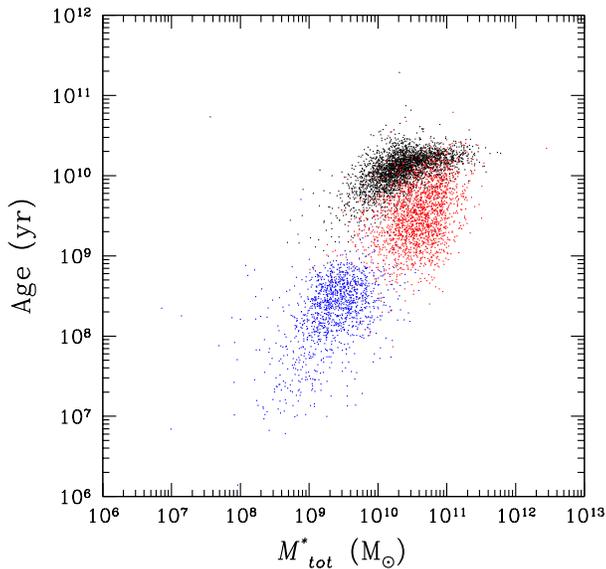}
\caption{Relationship between the age of the nuclear stellar population and the total stellar mass found in the SFG, S2G, and NG samples. Symbols are the same as in Fig.~8.}
\end{center}
\end{figure}
A look at the relationship between the stellar population ages in the circum-nuclear regions of galaxies and their total stellar masses, illustrated in Fig.~9, also reveals some further interesting effects. SFGs and S2Gs, indeed, lie on what appears to be a well defined sequence of increasing nuclear population age for larger total mass. The sequence has a high mass break and the range of masses, where AGN activity is observed, is also populated by NGs. The total stellar mass, therefore, increases regularly with age, through the star forming and nuclear activity stages. Later on, the stellar mass assembly is stopped and galaxies evolve as quiescent objects, dominated by old populations. Before drawing any conclusions, however, we have to consider the possible biases, introduced by our sample selection criteria. In particular, the choice of a $S / N$ limiting threshold reflects in the selection of more luminous objects. SFGs are not particularly prone to this problem, because the high luminosity of their young stellar populations enables us to detect objects with fairly small masses, but for S2Gs and NGs this could be a real issue. If we look at the mass distribution of the SDSS line emitting sources, which, in the considered redshift range, fall in the AGN region of the $O_{123}$ diagram, however, it turns out that a $S / N > 10$ threshold is not systematically excluding low mass objects from the sample, so that the observed mass range for S2Gs is a real feature in the considered volume. The NG sample, on the other hand, experiences a quite different situation. In this case, we are not able to include very low mass quiescent objects, but the issue is not particularly critical for our interests, because the artificial cut-off falls below the transition region between star forming and active galaxies. Moreover, as we noted in the analysis of the average spectral properties, in order for the difference between the spectra of S2Gs and NGs to have an appreciable physical meaning, we need that the NG sample covers a similar mass range, with respect to the S2Gs. Therefore, the adopted limiting constraints do not affect the S2G distribution, though we may expect that the NG population extends well below the detected lower end of the S2G mass range. Further investigation on the properties of low mass NGs would represent a useful extension of our work, but the task would require extensive spectral modeling of large numbers of spectra, which is beyond the scope of our current analysis.

\begin{figure}
\begin{center}
\includegraphics[width=8.4cm]{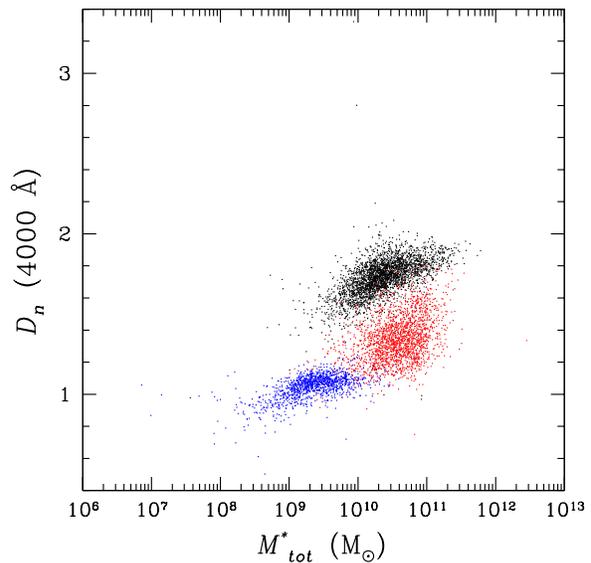}
\caption{$D_n(4000\, {\rm \AA})$ spectral index plotted against the total estimated mass of SFGs, S2Gs, and NGs. Symbols are the same as in Fig.~8.}
\end{center}
\end{figure}
The differences that we appreciated on the basis of our averaged analysis can be better understood by looking at the diagram plotted in Fig.~10, where we show the distribution of the $D_n\,$(4000~\AA) spectral index \citep[defined as in][]{Kauffmann03a} as a function of the total stellar mass. The region covered by SFGs and S2Gs represents a well defined sequence of regular growth in the index, with increasing host galaxy stellar mass. Quiescent galaxies, instead, are well separated from this sequence. The situation is fully consistent with a smooth transition of gradually increasing stellar ages in a generation of stars formed in pristine gas, as soon as we move from low mass SFGs to high mass SFGs, all the way across S2Gs. The transition region can be identified within a range of total stellar masses of $9.8 \leq \log (M_* / M_\odot) \leq 10.3$. The overlapping mass range may represent a region where star forming activity is most likely going to start an interplay with the AGN. Depending on the interaction between these processes, the galaxy may either stop its stellar mass assembly, eventually evolving as a low mass quiescent system, or form new generations of stars, like the high metallicity populations detected in some S2G spectra of our sample, to become a high mass system. The difference between the two evolutionary possibilities probably resides in the power of the nuclear source and the way it affects ISM in the surrounding space.

\section{Conclusions}
As we subdivide our observational sample in objects with different types of nuclear activity, taking into account Seyfert 2, star forming nuclei and normal galaxies, we observe that the circum-nuclear environments are characterized by appreciably different properties. Objects featuring ionized gas in their nuclear spectra show distinct ISM chemical compositions, characteristic of pristine gas in SFGs and of evolved medium in S2Gs. The stellar populations, however, do not reflect such a dichotomy, being almost everywhere consistent with star formation powered by pristine gas. Another striking property is the appreciably tight relationship, which holds between the age of stars and the global stellar mass of galaxies. A relationship between the age of stars and the stellar mass observed in galaxies is expected, if more massive systems are also forming more quickly. Since the circum-nuclear regions of active galaxies appear to be more likely explained as the result of a single star formation event, rather than as the combination of several loops, our analysis suggests that nuclear activity takes place after the bulk of stellar mass assembly has occurred. Because of its role in controlling the star formation time-scale, the observed stellar mass is likely a key parameter in determining whether an AGN can be triggered and if star formation can affect the circum-nuclear regions further on.

The average trends that we identified in our sample actually mark an important distinction in the environment of individual galaxies, whose nuclei are affected by different degrees of activity. The properties observed within a wide sample of galaxies, including normal objects, as well as star forming and active ones, are consistent with the expectation that star formation and nuclear activity should be actually related in the evolution of galaxies. The connection can be largely supported on the basis of sound theoretical grounds, but it is still missing a conclusive observational confirmation. In our analysis we identified several properties of the selected galaxies, which may lead to a more precise investigation of the problem and, ultimately, to a clearer view of the general picture.

The main results, which we are able to point out on the basis of our data, appear as follows:
\begin{enumerate}
\item there is a range of galaxy masses ($9.8 \leq \log (M_* / M_\odot) \leq 10.3$) where SFGs and AGNs are characterized by circum-nuclear stellar populations of approximately the same age and similar chemical abundances in the stellar component;
\item objects in this range are the oldest SFGs and the youngest AGNs (in terms of stellar population);
\item although the metallicity of stars in SFGs within this mass range are indistinguishable from those of lower mass objects and from most of the S2Gs, there are some S2Gs showing evidence for further evolution.
\end{enumerate}

The metallicity of the stars seems to be independent from the mass of the whole galaxy and from the age of the circum-nuclear stellar population, as it is the case for the gaseous component. This suggests that we are observing different stages of the evolution of the same star formation process. Therefore, the circum-nuclear regions of SFGs with mass $9.8 \leq \log (M_* / M_\odot) \leq 10.3$, which are clearly in a more advanced evolutionary stage than lower mass SFGs, seem to be the most promising candidate precursors of AGNs, if there is any causal link between the two phenomena. These results are in good agreement with the conclusions reported in the high resolution studies of \citet{Davies07}, who, investigating the nuclear environment on a much smaller scale, in a few particular cases, were able to detect co-existing star formation and nuclear activity and to estimate the typical time-scales connecting the two processes. Further evidence for a connection between star formation and nuclear activity from a much more similar dataset was also achieved by \citet{Shavinski07}, in their analysis of the role of AGN feedback in the evolution of galaxies. Looking at a sample of early type objects in the redshift range $0.05 \leq z \leq 0.1$, they concluded that AGN feedback is a key process in the evolution of a galaxy through its stages of star formation, nuclear activity and quiescent existence. They also conclude that the AGN feedback process is likely to smoothly affect the star formation rate of a galaxy, during time, and that more massive objects experienced this transition earlier.

\vspace{14pt}
\begin{center}
{\bf ACKNOWLEDGMENTS}
\end{center}
The authors gratefully thank the anonymous referee for discussion and suggestions leading to the improvement of this work. Data for this analysis come from the Sloan Digital Sky Survey. Funding for the SDSS and SDSS-II has been provided by the Alfred P. Sloan Foundation, the Participating Institutions, the National Science Foundation, the U.S. Department of Energy, the National Aeronautics and Space Administration, the Japanese Monbukagakusho, the Max Planck Society, and the Higher Education Funding Council for England. The SDSS Web Site is http://www.sdss.org/.

The SDSS is managed by the Astrophysical Research Consortium for the Participating Institutions. The Participating Institutions are the American Museum of Natural His- tory, Astrophysical Institute Potsdam, University of Basel, University of Cambridge, Case Western Reserve University, University of Chicago, Drexel University, Fermilab, the Institute for Advanced Study, the Japan Participation Group, Johns Hopkins University, the Joint Institute for Nuclear Astrophysics, the Kavli Institute for Particle Astrophysics and Cosmology, the Korean Scientist Group, the Chinese Academy of Sciences (LAMOST), Los Alamos National Laboratory, the Max-Planck-Institute for Astronomy (MPIA), the Max-Planck-Institute for Astrophysics (MPA), New Mexico State University, Ohio State University, University of Pittsburgh, University of Portsmouth, Princeton University, the United States Naval Observatory, and the University of Washington.

\end{document}